# Observation of Mixed Alkali Like Behaviour by Fluorine Ion in Mixed Alkali Oxyfluro Vanadate Glasses: Analysis from Conductivity Measurements.

Gajanan V Honnavar[1,2, a], Vaibhav Varade[3], Parul Goel[4] and K P Ramesh[5]

[1] Department of Physics, College of Science, Bahir Dar University, Bahir Dar P.O.-79, Ethiopia
[2] On sabbatical leave from Department of Physics, PES Institute of Technology – Bangalore South Campus, Near Electronic City, Hosur Road, Bangalore – 560 100, India
[3] Department of Chemical Physics, Weizmann Institute of Science, Rehovot - 76100, Israel

[a] Corresponding author: gajanan.honnavar@gmail.com

**Abstract.** In this communication we report the fluorine ion dynamics in mixed alkali oxyfluro vanadate glasses. We have fabricated glasses with general formula, $40V_2O_5 - 30BaF_2 - (30-x) LiF - xRbF$ where x = 0,5, 10,15,20,25 and 30. We have measured the electrical conductivity using impedance spectroscopy technique Room temperature conductivity falls to 5 orders of magnitude from its single alkali values at 33 mol% of rubidium concentration. We have also estimated the distance between similar mobile ions using the density values. Assuming this distance as the hopping distance between the similar ions we have estimated the anionic (Fluorine ion in our case) conductivity. It is observed that the fluorine ion dynamics mimics the mixed alkali effect and scales as the onset frequency $f_0$.

## 1. Introduction

Electrical transport is a long standing but important topic of research in glass science [1,2]. Transport phenomena in glasses are usually attributed to ion hopping, but electronic / polaronic processes are also observed in glasses containing transition metal ions [3,4]. Since glasses are highly resistive, impedance technique can provide a better insight in to the transport mechanisms and therefore is of importance. Moreover, a temperature dependent study of dielectric behaviour in such disordered systems can provide useful insight about carrier relaxation mechanism.

One of the interesting observation as far as dynamics of the mobile ions are concerned is mixed mobile ion effect (MMI) in glasses. This was first observed in glasses with alkali ions as the mobile carriers and hence it was called as Mixed Alkali Effect (MAE) [5]. Until recent past, understanding the non-linear behaviour of the mobile ions in mixed alkali glasses were a challenge. Now the scientific community has reached some consensuses about the origins of MAE [6]. This is attributed to the structural aspect of the mixed alkali glasses [7]. Thus there is renewed interest in studying MAE from the point of view of structural aspect.

Vanadate glasses are important family of glasses as they find applications in memory and switching devices. Vanadium shows a rich oxidation state ($V^{2+}$, $V^{3+}$, $V^{4+}$ and $V^{5+}$), which can be exploited to be used as battery materials. In glasses, Vanadium exhibits $V^{4+}$ and $V^{5+}$ states that are responsible for the electrical conduction via polaron hopping [8] and interesting structural aspect. [9].

In our previous communications we have studied mixed alkali oxyfluoro vanadate glasses to understand the structural origin of MAE using Raman spectroscopy [10] and Electron Paramagnetic Resonance (EPR) [9]. We report here the electrical conductivity analysis of the electrical impedance data of the above glass samples from the point of view of hopping ionic conductivity and calculate the contribution of fluorine ion conductivity and its dynamics. It is observed that the fluorine ion shows conductivity behaviour like that of mixed alkali.

## 2. Experiment and Results:
## 2.1 Experimental Details:
Glasses with general formula $40V_2O_5 - 30BaF_2 - (30-x) LiF - xRbF$ where x = 0,5, 10,15,20,25 and 30 are prepared via melt quenching technique. The results of basic characterization like XRD, DSC, Density are discussed in detail in one of our previous communication [9].

The electrical characterization of these samples were carried out using Agilent Precision Impedance Analyzer 4194A in the frequency range 100 Hz to 10 MHz. Complex conductivity $\hat{\sigma} = 2\pi f \varepsilon_0 \hat{Z} k$, where $\hat{Z}(f)$ - complex impedance measured from the impedance analyser, $k$ – cell constant given by $d/A$ ( $d$ is the thickness and $A$ is the area of cross section of the sample)[11] was calculated for each sample and for all the temperature range from 140 °C to 300 °C. We report here only the analysis of the frequency independent DC part of the conductivity data. The technical details of the impedance analyser, temperature controller, the procedure followed and the electrical modulus analysis of the data will be detailed out in a separate communication [12].

## 2.2 Results:
The real part of complex conductivity $\hat{\sigma}(f)$ reflect the dynamics of the ion transport and is known to follow the power law behaviour[13]:

$$\sigma' = \sigma_0 \left[1 + \left(\frac{f}{f_0}\right)^p\right] \quad (1)$$

The first term in Eq. (1) represents the constant dc plateau and the second term describes the dispersive region, which sets in approximately at the crossover frequency onset $f_0$; $p$ is the power law exponent. The product of $\sigma_0$ and temperature $T$ follows Arrhenius behaviour below glass transition temperature given as:

$$\sigma_0 T = A_0 \exp\left(-\frac{E_0}{k_B T}\right) \quad (2)$$

$A_0$ is a pre-factor, $E_0$ is the activation energy for ion hopping and physically signifies the energy of the potential barrier offered by the lattice for mobile ions to overcome and $k_B$ is Boltzman's constant.

The calculated complex conductivity values were plotted against frequency (shown in Fig. 1(a)) and were fitted to Eq. (1). Extrapolation of $\sigma'(f)$ to low frequencies yields the dc conductivity, $\sigma_0$. The values of crossover frequency $f_0$ and the power law exponent $p$ were extracted from the fit. The value of $p$ lies in the range 0.6 - 0.8. Table 1 summarizes these parameters from the fitting of Eq. (1) to the conductivity spectra. It also includes the estimated conductivity at room temperature from the Arrhenius plots of Fig.1(b). The Arrhenius plots of all the glass samples are shown. The slope of each straight line corresponds to dc activation energy for ion hopping in respective glass sample. The room temperature dc conductivity values were obtained by extrapolating the corresponding Arrhenius plots of glasses to low temperature region [14]. Fig 1(c) and (d) shows the variation of activation energy and dc conductivity with molar fraction of Rubidium ions respectively. In Fig 1(d) one can clearly observe that room temperature dc conductivity shows a drastic decrease in the value at 33 mol% of Rubidium content. The decrease is quite high of the order of $10^5$ ($\Omega^{-1}cm^{-1}$). This is attributed to the MAE. In the same plot, two high temperature dc conductivity values are plotted for comparison. The dip in the conductivity value is not as pronounced as that of room temperature. It is also interesting to observe that the conductivity minima shift to the higher mol% of Rubidium content as the temperature increases.

## 3. Analysis and Discussion:
The dc conductivity in semiconducting glasses like vanadates and tellurites, are usually interpreted on the basis of polaronic [4] or ionic conductivity. In our study we eliminate out the contribution from the polaronic conductivity due to following reason. The time scale of the ionic vibrations and electronic motions are of the order of $10^{-12}$ s and $10^{-14}$ s respectively, these processes can be regarded as infinitely fast within the time scale of a typical ac impedance experiment (1µs to 1s). Therefore, the σ' spectra and hence $\sigma_0$ is mainly governed by the ion transport properties of the glasses.

To connect the dc conductivity to the microscopic processes, one brings in the relation between mean square displacement between the mobile ions and the conductivity, which was established using linear response theory [11]. The activation energy for the conductivity is almost entirely due to the energy barrier between two sites. We follow Karlsson et. al., [14], in which it was shown that the temperature dependency of $\sigma_0$ is essentially derived from the hopping frequency, $f_h$, between two sites. The hopping frequency is proportional to the onset frequency, $f_0$. [11]. From the above arguments, it is clear that for a given temperature:

$$\sigma_{0,T} \propto f_h \propto f_0$$

Where $\sigma_{0,T}$, represents the dc conductivity at a given temperature T.

In Fig. 2(a), we plot the dc conductivity data of 220 °C. We observe near correspondence between dc conductivity at a given temperature $\sigma_{0,T}$ and onset frequency, $f_0$ which provides support for our assumption that $\sigma_{0,T}$ scales with $f_0$. Number density of mobile ion scaling with temperature was not observed in literatures so it was not considered here. There is variation in the concentration of mobile cations (lithium and rubidium). This is evident from Table 1 in which we have listed the concentrations of alkali ions calculated from the density data. From these data, one can calculate the distance between similar alkali ions [15]. This distance may be assumed the hopping length for alkali ions of same type [16]. Thus, the conductivity due to a particular ion can be estimated using the formula:

$$\sigma_{0,T} = \frac{nq^2 a^2}{6k_B T} \times f_0 \tag{3}$$

'$n$' is the number density of the mobile ions, '$q$' is the elementary charge and '$a$' is the hopping distance. Other symbols have their usual meaning. It is observed that the presence of fluorine ions in the glass enhances the electrical conductivity from its oxide counterpart [17]. This may be due to the increased iconicity of the glass by the inclusion of fluorine ion. The fluorine ions rupture the glass network and introduce more non-bridging oxygen [15]. The conductivity calculated from Eq. (3) using concentration of fluorine ions and the distance between them as the hopping distance, shows an interesting trend. This is shown in Fig.2(b). The fluorine ion conductivity scales exactly as onset frequency, $f_0$. This again confirms our assumption that conductivity $\sigma_{0,T}$ scales with $f_0$.

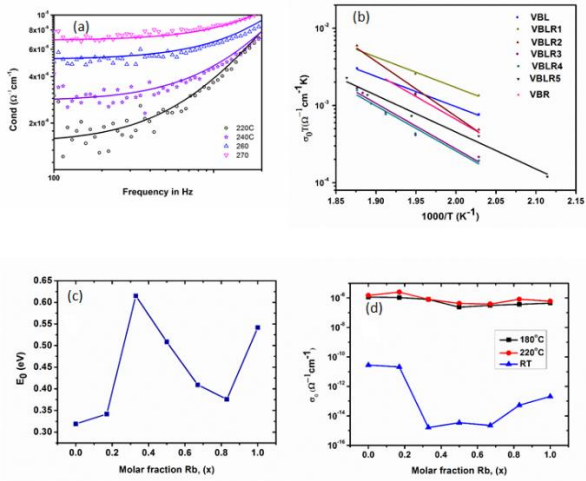
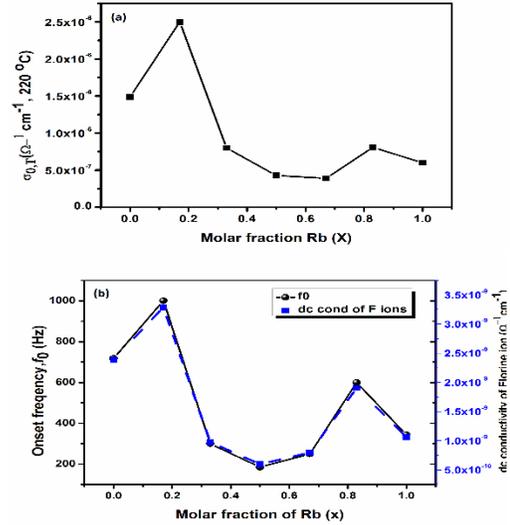

**FIGURE 1.** (a): Representative graph of $\sigma'$ at different temperatures. The solid line is the fit of Equation 1. (b): Arrhenius plots of $\sigma_0 T$ for all the glass samples. The solid lines are the fit to Eq. (2). (c): The activation energy for ion hopping calculated from the slope of (b). (c): Room temperature (RT) conductivity extracted from (b) and the conductivity at 180 °C and 220 °C for comparison.

**FIGURE 2.** (a): The conductivity of all the glass samples as a function of composition at 220 °C. The molar fraction of Rb is given in Table 1. (b): Scaling of fluorine ion conductivity and onset frequency f0, extracted from the fit to Eq. (1).

**TABLE 1:** $\sigma_0$ AND $n$. SEE TEXT FOR THE MEANING OF THE SYMBOLS. THE ERROR IN $\sigma_0$ VARIES BETWEEN 5 TO 20 %. $n$ IS CALCULATED USING DENSITY DATA [9].

| Batch Code | $\frac{Rb}{Li+Rb}$ | $\sigma_0$ at 220 ºC in $\mu\Omega^{-1}$ cm$^{-1}$ | Number of Li ions in $10^{18}$ ions / cc | Number of Rb ions in $10^{18}$ ions / cc | Number of F ions in $10^{20}$ ions / cc | Average distance between F ions in $10^{-7}$ cm | O/F |
|---|---|---|---|---|---|---|---|
| VBL | 0 | 1.49 | 49.95 | 0 | 1.49 | 1.88 | 2.22 |
| VBLR1 | 0.17 | 2.5 | 40.04 | 8.01 | 1.44 | 1.91 | 2.22 |
| VBLR2 | 0.33 | 0.8 | 30.87 | 15.43 | 1.39 | 1.93 | 2.22 |
| VBLR3 | 0.5 | 0.43 | 22.92 | 22.92 | 1.38 | 1.94 | 2.22 |
| VBLR4 | 0.67 | 0.39 | 14.41 | 28.88 | 1.29 | 1.97 | 2.22 |
| VBLR5 | 0.83 | 0.81 | 7.22 | 36.11 | 1.29 | 1.97 | 2.22 |
| VBR | 1 | 0.6 | 0 | 39.81 | 1.19 | 2.03 | 2.22 |

## 4. Conclusions:

We have fabricated mixed alkali oxyfluoro vanadate glasses with lithium and rubidium as the mobile ions. We have measured the electrical conductivity using impedance spectroscopy technique. We have calculated electrical conductivity at room temperature, $\sigma_{0,RT}$ and activation energy for ion hopping, $E_0$ using the conductivity data. Both parameters show mixed alkali effect. Room temperature conductivity falls to 5 orders of magnitude from its single alkali values at 33 mol% of rubidium concentration. We have also estimated the distance between similar mobile ions using the density values. Assuming this distance as the hopping distance between the similar ions we have estimated the anionic (Fluorine ion in our case) conductivity. It is observed that the fluorine ion dynamics mimics the mixed alkali effect and scales as the onset frequency $f_0$.